\pdfoutput=1

\documentclass[11pt]{article}

\usepackage[final]{acl}

\usepackage{times}
\usepackage{latexsym}
\usepackage[T1]{fontenc}
\usepackage[utf8]{inputenc}
\usepackage{microtype}
\usepackage{inconsolata}

\usepackage{color,xcolor}
\usepackage{epsfig}
\usepackage{graphicx}
\usepackage{lipsum}
\usepackage{array}
\usepackage{booktabs}
\usepackage{colortbl}
\usepackage{multirow}
\usepackage{float}
\usepackage{caption}
\usepackage{adjustbox}
\usepackage{subfigure}
\usepackage{footnote}
\makesavenoteenv{tabular}
\makesavenoteenv{table}
\usepackage{makecell}
\usepackage{tablefootnote}

\usepackage{amsmath,amsfonts,amssymb,bm}
\usepackage[super]{nth}

\usepackage{changepage}
\usepackage{extramarks}
\usepackage{lastpage}
\usepackage{setspace}
\usepackage{soul}
\usepackage{xspace}

\usepackage{url}
\usepackage{hyperref}
\hypersetup{colorlinks=True, urlcolor=black}

\usepackage[ruled,vlined]{algorithm2e}
\usepackage{enumitem}
\usepackage{verbatim}
\usepackage{pifont}
\usepackage[acronyms]{glossaries}
\glsdisablehyper

\title{GigaSpeech 2: An Evolving, Large-Scale and Multi-domain ASR Corpus for Low-Resource Languages with Automated Crawling, Transcription and Refinement}

\author{
\textbf{Yifan Yang\textsuperscript{1}},
\textbf{Zheshu Song\textsuperscript{1}},
\textbf{Jianheng Zhuo\textsuperscript{1}},
\textbf{Mingyu Cui\textsuperscript{3}}
\\
\textbf{Jinpeng Li\textsuperscript{4}},
\textbf{Bo Yang\textsuperscript{2}},
\textbf{Yexing Du\textsuperscript{2,6}},
\textbf{Ziyang Ma\textsuperscript{1}},
\textbf{Xunying Liu\textsuperscript{3}},
\textbf{Ziyuan Wang\textsuperscript{7}}
\\
\textbf{Ke Li\textsuperscript{8}}, 
\textbf{Shuai Fan\textsuperscript{9}}, 
\textbf{Kai Yu\textsuperscript{1,9}},
\textbf{Wei-Qiang Zhang\textsuperscript{4,11}}, 
\textbf{Guoguo Chen\textsuperscript{10,11}},
\textbf{Xie Chen\textsuperscript{1,5,11}\thanks{Corresponding author. Email:  \href{mailto:chenxie95@sjtu.edu.cn}{chenxie95@sjtu.edu.cn}}}
\\
\textsuperscript{1}X-LANCE Lab, SCS, MoE Key Lab of Artificial Intelligence, Shanghai Jiao Tong University \\
\textsuperscript{2}Pengcheng Laboratory
\textsuperscript{3}The Chinese University of Hong Kong
\textsuperscript{4}Dept EE, Tsinghua University
\textsuperscript{5}SII \\
\textsuperscript{6}Harbin Institute of Technology
\textsuperscript{7}Birch AI
\textsuperscript{8}Dataocean AI
\textsuperscript{9}AISpeech Ltd
\textsuperscript{10}Seasalt AI Inc
\textsuperscript{11}SpeechColab
}

\begin{document}

\maketitle

\begin{abstract}
The evolution of speech technology has been spurred by the rapid increase in dataset sizes. Traditional speech models generally depend on a large amount of labeled training data, which is scarce for low-resource languages. This paper presents GigaSpeech 2, a large-scale, multi-domain, multilingual speech recognition corpus. It is designed for low-resource languages and does not rely on paired speech and text data. GigaSpeech 2 comprises about 30,000 hours of automatically transcribed speech, including Thai, Indonesian, and Vietnamese, gathered from unlabeled YouTube videos. We also introduce an automated pipeline for data crawling, transcription, and label refinement. Specifically, this pipeline involves Whisper for initial transcription, MMS for forced alignment, and multi-dimensional filtering for data quality assurance. A modified Noisy Student Training is developed to further refine flawed pseudo labels iteratively, thereby enhancing model performance. Experimental results on our manually transcribed evaluation set and two public test sets from Common Voice and FLEURS confirm our corpus's high quality and broad applicability. Notably, ASR models trained on GigaSpeech 2 can reduce the word error rate for Thai, Indonesian, and Vietnamese on our challenging and realistic YouTube test set by 25\% to 40\% compared to Whisper large-v3, with merely 10\% model parameters. Furthermore, our ASR models trained on GigaSpeech 2 yield superior performance compared to commercial services. We hope that our newly introduced corpus and pipeline will open a new avenue for low-resource speech recognition and significantly facilitate research in this area.
\end{abstract}
\section{Introduction}
In recent years, the scaling of model parameters and data size has prevailed and proven effective in a range of areas, including language~\cite{openai_scalinglaw, hoffmann_scalinglaw}, vision~\cite{betker2023improving, dehghani2023scaling}, as well as speech processing~\cite{mms, usm, whisper}.
Consequently, pursuing superior AI models is now closely associated with expanding model size and leveraging larger, high-quality datasets. In the realm of Automatic Speech Recognition (ASR), several large-scale open-source labeled speech datasets~\cite{gigaspeech, libriheavy, wenetspeech, peoplespeech, mls, commonvoice} have been proposed.
However, these extensive datasets are only available for several mainstream languages, such as English and Mandarin, hindering speech recognition development for low-resource languages.
Moreover, traditional ASR corpus~\cite{commonvoice, fleurs, aishell1, aishell2} construction relies heavily on human-labeled speech data, making it time-consuming and a major bottleneck in the fast-paced AI industry. Reducing dependence on vast labeled data is crucial when expanding to new languages and domains.
YODAS~\cite{yodas} attempts to address this issue by building multilingual datasets via scraping audio and transcriptions from YouTube. However, neither manual nor automatic subtitles accurately reflect the speech content, resulting in unguaranteed quality.

With this perspective in mind, we propose a new paradigm for constructing large-scale ASR datasets, focusing solely on audio content irrespective of the existence or quality of corresponding text pairs. This approach leverages the gigantic amount of unlabeled audio data, bypassing the constraints of scarce paired data.
We introduce GigaSpeech 2, an evolving\footnote{The term ``evolving'' continues the naming convention used by GigaSpeech.}, large-scale, multi-domain, multilingual ASR corpus for low-resource Southeast Asian languages. \textit{GigaSpeech 2 raw} comprises about 30,000 hours of automatically transcribed speech, across Thai, Indonesian, and Vietnamese. \textit{GigaSpeech 2 refined} consists of 10,000 hours of Thai, 6,000 hours each for Indonesian and Vietnamese.
To achieve this, an automated pipeline is developed for data crawling, transcription, and filtering.
Furthermore, a modified Noisy Student Training (NST)~\cite{nst_cv} method is proposed to refine labels from flawed data iteratively.
Through comprehensive evaluations, ASR models trained on \textit{GigaSpeech 2 refined} can reduce the word error rate for Thai, Indonesian, and Vietnamese on our YouTube test set by 25\% to 40\% compared to the powerful Whisper large-v3 model, with merely 10\% model parameters. 

Our contributions can be summarized as follows:
\begin{itemize}[leftmargin=*,noitemsep]
    \item We release GigaSpeech 2 with two versions: \textit{GigaSpeech 2 raw} comprises about 30,000 hours of automatically transcribed speech across Thai, Indonesian, and Vietnamese. \textit{GigaSpeech 2 refined} consists of 10,000 hours of Thai, 6,000 hours each for Indonesian and Vietnamese.

    \item We develop an automated pipeline for data crawling, transcription, and label refinement, enabling the creation of large-scale speech datasets without reliance on labeled data.

    \item We propose a modified NST method to iteratively refine flawed pseudo labels. Our modified NST performs scaling, relabeling, and filtering data within each iteration, significantly improving final data quality.

    \item We release a series of challenging and realistic speech recognition test sets, including Thai, Indonesian, and Vietnamese. Compared to previous public test sets, GigaSpeech 2 test sets more realistically reflect speech recognition scenarios and mirror the real performance of an ASR system for low-resource languages. 
    
    \item Experimental results on our challenging GigaSpeech 2 test sets, as well as other competitive public test sets including Common Voice and FLEURS, demonstrate the superiority of the ASR models trained on GigaSpeech 2 over several competitive baselines, including Whisper large-v3 and commercial services.
\end{itemize}

\section{Related Work}
\begin{table*}[t]
  \centering
  \small
  \caption{Comparison of data size between GigaSpeech 2 and other common public multilingual speech datasets on Thai (th), Indonesian (id), and Vietnamese (vi).}
  \label{tab:related_dataset}
  \renewcommand{\arraystretch}{1.1}
  \renewcommand\tabcolsep{1.5pt}
  \resizebox{\linewidth}{!}{
    \begin{tabular}{ccccccc}
    \toprule[1pt]
    \multirow{2}{*}{\textbf{Dataset}} & \multirow{2}{*}{\textbf{Language}} & \multirow{2}{*}{\makecell[c]{\textbf{\# Hours} \\ \textbf{(h)}}} & \multirow{2}{*}{\textbf{Domain}} & \multirow{2}{*}{\textbf{Speech Type}} & \multirow{2}{*}{\textbf{Labeled}} & \multirow{2}{*}{\textbf{Label Type}} \\
    \\ \hline
    \multirow{3}{*}{Common Voice~\cite{commonvoice}}     & th & 172.0   & \multirow{3}{*}{Open domain}  & \multirow{3}{*}{Read}        & \multirow{3}{*}{Yes} & \multirow{3}{*}{Manual} \\ 
                                                         & id & 28.0    &                               &                              &                      & \\
                                                         & vi & 6.0     &                               &                              &                      & \\ \hline
    \multirow{3}{*}{FLEURS~\cite{fleurs}}                & th & 13.3    & \multirow{3}{*}{Wikipedia}    & \multirow{3}{*}{Read}        & \multirow{3}{*}{Yes} & \multirow{3}{*}{Manual} \\
                                                         & id & 12.6    &                               &                              & \\ 
                                                         & vi & 13.3    &                               &                              & \\ \hline
    \multirow{3}{*}{VoxLingua107~\cite{voxlingua107}}    & th & 61.0    & \multirow{3}{*}{YouTube}      & \multirow{3}{*}{Spontaneous} & \multirow{3}{*}{No}  & \multirow{3}{*}{-} \\ 
                                                         & id & 40.0    &                               &                              & \\
                                                         & vi & 64.0    &                               &                              & \\ \hline
    \multirow{3}{*}{CMU Wilderness~\cite{cmuwilderness}} & th & 15.6    & \multirow{3}{*}{Religion}     & \multirow{3}{*}{Read}        & \multirow{3}{*}{Yes} & \multirow{3}{*}{Manual} \\ 
                                                         & id & 70.9    &                               &                              & \\
                                                         & vi & 9.2     &                               &                              & \\ \hline
    BABEL~\cite{babel}                                   & vi & 87.1    & Conversation                  & Spontaneous                  & Yes                  & Manual \\ \hline
    VietMed~\cite{vietmed}                               & vi & 16.0    & Medical                       & Spontaneous                  & Yes                  & Manual \\ \hline
    Thai Dialect Corpus~\cite{thaidialectcorpus}         & th & 840.0   & Open domain                   & Read                         & Yes                  & Manual \\ \hline
    TITML-IDN~\cite{titml-idn}                           & id & 14.5    & News                          & Read                         & Yes                  & Manual \\ \hline
    MEDISCO~\cite{medisco}                               & id & 10.0    & Medical                       & Read                         & Yes                  & Manual \\ \hline
    \multirow{3}{*}{YODAS manual~\cite{yodas}}           & th & 497.1   & \multirow{3}{*}{YouTube}      & \multirow{3}{*}{Spontaneous} & \multirow{3}{*}{Yes} & \multirow{3}{*}{Manual} \\ 
                                                         & id & 1420.1  &                               &                              & \\
                                                         & vi & 779.9   &                               &                              & \\ \hline
    \multirow{3}{*}{YODAS automatic~\cite{yodas}}        & th & 1.9     & \multirow{3}{*}{YouTube}      & \multirow{3}{*}{Spontaneous} & \multirow{3}{*}{Yes} & \multirow{3}{*}{Pseudo} \\ 
                                                         & id & 8463.6  &                               &                              & \\
                                                         & vi & 9203.1  &                               &                              & \\
    \midrule[1pt]
    \multirow{3}{*}{\textit{GigaSpeech 2 raw}}           & th & 12901.8 & \multirow{3}{*}{YouTube}      & \multirow{3}{*}{Spontaneous} & \multirow{3}{*}{Yes} & \multirow{3}{*}{Pseudo} \\ 
                                                         & id & 8112.9 &                               &                              & \\
                                                         & vi & 7324.0  &                               &                              & \\ \hline
    \multirow{3}{*}{\textit{GigaSpeech 2 refined}}       & th & 10262.0 & \multirow{3}{*}{YouTube}      & \multirow{3}{*}{Spontaneous} & \multirow{3}{*}{Yes} & \multirow{3}{*}{Pseudo} \\ 
                                                         & id & 5714.0  &                               &                              & \\
                                                         & vi & 6039.0  &                               &                              & \\ 
\bottomrule[1pt]
\end{tabular}}
\end{table*}

\noindent\textbf{Multilingual Low-Resource Speech Datasets}\quad
Several publicly available multilingual speech datasets have emerged for low-resource languages. BABEL~\cite{babel}, a pioneering dataset, includes conversational telephone data in 17 African and Asian languages. Common Voice~\cite{commonvoice} offers 19,000 hours of validated recordings in over 100 languages. FLEURS~\cite{fleurs} covers 102 languages with 12 hours of supervised data per language. CMU Wilderness~\cite{cmuwilderness} provides 20 hours of New Testament data for over 700 languages. VoxLingua107~\cite{voxlingua107} contains 6,628 hours of unlabeled YouTube data across 107 languages.
However, most public multilingual speech datasets focus on high-resource languages, leaving low-resource languages with limited annotated speech data. As detailed in Table~\ref{tab:related_dataset}, the available open-source data for Thai, Indonesian, and Vietnamese is scarce. In contrast, industry-utilized speech models like Whisper~\cite{whisper}, MMS~\cite{mms}, Google USM~\cite{usm}, and Universal-1~\cite{universal-1} are trained on massive industrial-grade datasets, the details of which remain undisclosed.
To resolve the problem, YODAS~\cite{yodas} attempts to crawl audio from YouTube, but neither manual nor automatic subtitles accurately reflect the speech content, resulting in unguaranteed quality.
Moreover, widely used evaluation benchmarks for low-resource languages~\cite{commonvoice,fleurs} only consist of read speech, which is relatively clean and mismatched with real-world speech data.

\noindent\textbf{Multilingual Automatic Speech Recognition}\quad
As the demand for communication between people worldwide grows, many works~\cite{whisper,usm,mms,multilingual1,multilingual2,multilingual3,multilingual4,multilingual5,multilingual6,multilingual7,xlsr} have shifted attention to multilingual speech recognition.
Whisper~\cite{whisper}, built on 680,000 hours of web data, supports 99 languages. Google USM~\cite{usm}, trained on YouTube audio, extends to 100+ languages. Massively Multilingual Speech (MMS)~\cite{mms}, trained on religion data, further scales to 1,107 languages.

\noindent\textbf{Noisy Student Training (NST)}\quad
NST~\cite{nst_cv,nst_speech,nst_speech2,nst_speech3,nst_speech4,nst_speech5,nst_speech6} is a self-training technique that leverages unlabeled data to enhance performance. Traditional NST methods start with training a teacher model on high-quality labeled data. Each student model then trains on both noisy-augmented labeled data and pseudo-labeled data generated by its teacher from the unlabeled data.
One study~\cite{nst_speech6} has explored using Character Error Rate (CER), calculated between pseudo-labeled data generated with and without language model, to perform data selection, suggesting a positive correlation between the CERs of different pseudo labels and their ground truth.
\begin{figure*}[t]
    \centering
    \includegraphics[width=\linewidth]{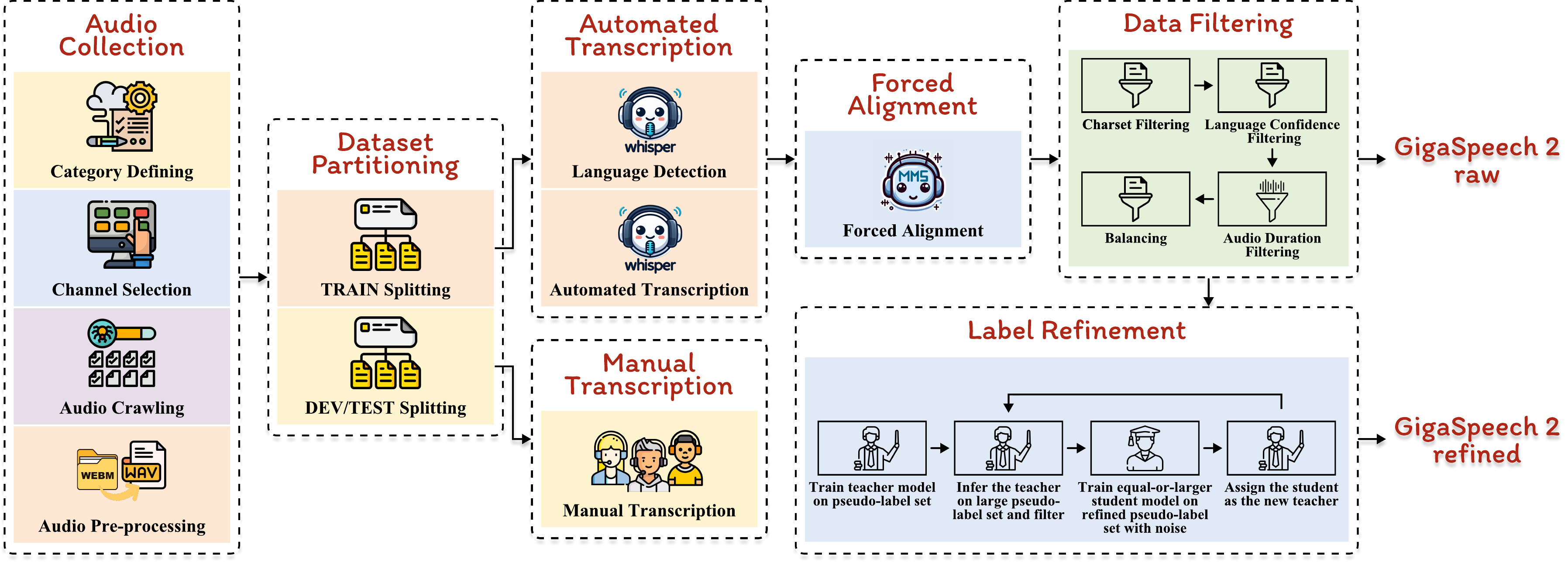}
    \caption{Automated construction pipeline of GigaSpeech 2, comprising (1) audio collection, (2) dataset partitioning, (3) automated transcription with Whisper, (4) forced alignment with TorchAudio, (5) transcription normalization, (6) data filtering, and (7) label refinement.}
    \label{fig:pipeline}
\end{figure*}

\section{Dataset Construction}

Our proposed automated construction pipeline is illustrated in Fig.~\ref{fig:pipeline}. Sec.~\ref{sec:build_gigaspeech2_raw} covers the stages involved in building \textit{GigaSpeech 2 raw} and Sec. ~\ref{sec:build_gigaspeech2_refined} further construct \textit{GigaSpeech 2 refined}.

\subsection{GigaSpeech 2 raw: Automated Crawling and Transcription}
\label{sec:build_gigaspeech2_raw}
\noindent\textbf{Audio Collection}\quad
Due to the scarcity of human-labeled data in low-resource languages, our dataset is collected with a focus solely on the audio content, irrespective of the existence or quality of corresponding text pairs. This strategy allows for leveraging a broader range of audio data.
Given the scarcity and uneven distribution of resources for low-resource languages, we strategically crawl videos from YouTube channels based on two key considerations.
First, prioritizing mainstream and popular channels helps ensure consistent domain characteristics and higher audio quality. Such content is widely viewed, and its creators are generally more mindful of ethical and legal considerations prior to publishing. Second, channels with huge differences in topics and content formats are less likely to have speaker overlap, which simplifies subsequent data partitioning.
The data collection process starts by manually defining categories of interest. The selected topics include Agriculture, Art, Business, Climate, Culture, Economics, Education, Entertainment, Health, History, Literature, Music, Politics, Relationships, Shopping, Society, Sport, Technology, and Travel. Alongside multiple topics, various content formats are also considered, including Audiobook, Commentary, Lecture, Monologue, Movie, News, Talk, and Vlog. This broad selection ensures the comprehensiveness of the dataset across multiple domains for research and analysis.
Once the list of YouTube channels is prepared, we use yt-dlp\footnote{\url{https://github.com/yt-dlp/yt-dlp}} toolkit to download all audio files in WebM format. These files are then converted to WAV format with a single channel and resampled at a 16 kHz sampling rate.

\noindent\textbf{Creating TRAIN/DEV/TEST Splits}\quad
To ensure no speaker overlap between the splits, we manually verify no speaker overlap between different channels and partition the data by allocating different YouTube channels to each subset. The dataset is divided into three distinct subsets: TRAIN, DEV, and TEST. The DEV and TEST sets each contain 10 hours and are manually transcribed by professionals, while the remainder is allocated to the TRAIN set. Table~\ref{tab:related_dataset} shows the amount of data across these three languages. Detailed analysis of GigaSpeech 2 is illustrated in Appendix~\ref{sec:detailed_analysis_gigaspeech2}.

\noindent\textbf{Transcription with Whisper}\quad
Whisper large-v3 model\footnote{\url{https://huggingface.co/openai/whisper-large-v3}} from OpenAI is used to transcribe audio files automatically. For each audio recording, a 30-second segment is selected from the middle to perform language detection by Whisper. Only audios that match the target languages are transcribed.

\noindent\textbf{Forced Alignment with TorchAudio}\quad
Although Whisper can generate timestamps, inspection reveals they are not precise enough. We resort to the model\footnote{\url{https://dl.fbaipublicfiles.com/mms/torchaudio/ctc_alignment_mling_uroman/model.pt}} from TorchAudio~\cite{torchaudio21} for forced alignment, which provides reliable alignment for noisy transcriptions, supports efficient processing on GPUs, and handles longer sequences more effectively \citep{mms}.

\noindent\textbf{Text Normalization}\quad
Text normalization on transcripts involves applying Normalization Form Compatibility Composition (NFKC), converting all characters to uppercase, removing punctuation, and mapping Arabic numerals to corresponding words in the respective languages.

\noindent\textbf{Multi-dimensional Filtering}\quad
A series of heuristic filtering rules across text and audio modalities are implemented to exclude relatively poor-quality samples.
1) Charset Filtering: Segments are retained if they only contain characters permitted by the charset of the respective language.
2) Language Confidence Filtering: The language identification (LID) model\footnote{\url{https://dl.fbaipublicfiles.com/fasttext/supervised-models/lid.176.bin}} from fastText~\citep{fasttext} is used to filter based on the estimated language confidence score, retaining only segments with confidence scores above a predetermined threshold. This method effectively eliminates meaningless and repetitive segments. Note that language identification based on audio has already been performed before transcription.
3) Audio Duration Filtering: Segments are filtered based on duration, with only those retained within the predetermined minimum and maximum duration thresholds.
4) Balancing: We carefully control the duplication of transcripts caused by channel-specific content while preserving natural linguistic patterns.

\subsection{GigaSpeech 2 refined: Iterative Label Refinement}
\label{sec:build_gigaspeech2_refined}
Some samples remain low quality due to inaccuracies in Whisper transcriptions and imprecise forced alignment boundaries. To address this, we develop a modified NST method. As illustrated in the bottom right corner of Fig.~\ref{fig:pipeline}, it begins by training a teacher model on a subset of flawed pseudo labels, iteratively expanding the training set, generating new pseudo labels, and filtering them. A student model, equal to or larger than the teacher, is trained on these refined pseudo labels and assigned as the new teacher. Unlike previous NST approaches that heavily rely on unchanged supervised data combined with additional unsupervised data, our method eliminates the need for any supervised data. Instead, we treat the flawed pseudo labels generated by Whisper as supervised data, refining all labels iteratively based on the Character Error Rate (CER) between those produced by Whisper and the teacher model. SpecAugment~\citep{specaugment}, Bypass~\citep{zipformer}, and feature mask~\citep{zipformer} introduce noise during each NST step. Bypass, a type of stochastic depth, learns channel-wise scalar weights to combine the module input and output. Feature mask performs dropout in the hidden dimension of the feedforward and convolution layer but shares across the time dimension. This deliberate noising enables the student model to learn consistency with the teacher model, which remains unaffected by noise when generating pseudo labels \citep{nst_cv}. This iterative process progressively enhances data quality.
Algo.~\ref{algo:nst} illustrates the workflow of our proposed iterative label refinement.

\begin{algorithm}[ht]
\caption{Iterative Label Refinement} 
\label{algo:nst} 
    \SetAlgoLined
    \KwIn{Pseudo-label set $\mathcal{P}$, Number of iterations $n$, Threshold $\tau$}
    \KwOut{Refined-label set $\mathcal{R}$}
    Divide $\mathcal{P}$ into $n$ splits $\mathcal{P}_1, \mathcal{P}_2, \ldots, \mathcal{P}_n$\;
    $\mathcal{R} \leftarrow \mathcal{P}_1$\;
    Train teacher model $\mathcal{M}_1$ on $\mathcal{R}$ with noise\;
    \For{$i \leftarrow 1$ \KwTo $n$} {
    $\mathcal{R} \leftarrow \varnothing$\;
        \eIf {$i$ == $1$ }{
            \tcp{Filter $\mathcal{P}_i$ by teacher model $\mathcal{M}_i$ with CER $\leq \tau$}
            $\mathcal{R} \leftarrow \{(x, y) \in \mathcal{P}_i \mid \mathrm{CER}(y,\mathcal{M}_{i}(x)) \leq \tau\}$\;
        } {
            \For{$j \leftarrow 1$ \KwTo $i$} {
                \tcp{Relabel $\mathcal{P}_j$ by teacher model $\mathcal{M}_i$ and filter with CER $\leq \tau$}
                $\mathcal{R}_{tmp} \leftarrow \{(x,\mathcal{M}_{i}(x)) \mid (x, y) \in \mathcal{P}_j, \mathrm{CER}(y,\mathcal{M}_{i}(x)) \leq \tau\}$\;
                $\mathcal{R} \leftarrow \mathcal{R} \cup \mathcal{R}_{tmp}$\;
            }
        }
        Train equal-or-larger student model $\mathcal{M}_{i+1}$ on $\mathcal{R}$ with noise and assign as new teacher\;
    }
    \Return $\mathcal{R}$\;
\end{algorithm}

\section{Experiments}
\subsection{ASR Model Training on GigaSpeech 2}
Our ASR systems are built on Zipformer Transducer~\cite{rnnt}. Two Zipformer~\cite{zipformer} variants, namely Zipformer-M and Zipformer-L, are employed for each NST iteration. Specific configurations are provided in Appendix~\ref{sec:config_zipformer}. During Noisy Student Training, 
SpecAugment~\cite{specaugment} is used as input noise while Bypass~\cite{zipformer} and feature mask~\cite{zipformer} are used as model noise.

\begin{table*}[t]
  \small
  \centering
  \caption{Comparison of ASR performance across different NST iterations on various evaluation sets, including GigaSpeech 2 DEV and TEST, Common Voice 17.0 TEST, and FLEURS TEST. Reported details include training set size (\#Hours), BPE vocabulary size (\#Vocab), model size (\#Params), CER for Thai, and WER for Indonesian and Vietnamese.}
  \label{tab:performance_nst}
  \renewcommand{\arraystretch}{1.15}
  \renewcommand\tabcolsep{8.0pt}
  \resizebox{\linewidth}{!}{
    \begin{tabular}{lccccccc}
    \toprule[1pt]
    \multirow{3}{*}{\makecell[c]{\textbf{NST} \\ \textbf{Iter}}} & \multirow{3}{*}{\makecell[c]{\textbf{\#Hours} \\ \textbf{(h)}}} & \multirow{3}{*}{\makecell[c]{\textbf{\#Vocab} \\ \;}} & \multirow{3}{*}{\makecell[c]{\textbf{\#Params} \\ \textbf{(M)}}} & \multicolumn{4}{c}{\textbf{CER / WER}} \\
    & & & & \multicolumn{2}{c}{\makecell[c]{GigaSpeech 2 \\ DEV \; TEST}} &  \makecell[c]{Common Voice \\ TEST} & \makecell[c]{FLEURS \\ TEST} \\ \hline
    \textbf{Thai} \\
    \hspace{1em} 1 & 4378  & 500  & 65.5 & 12.14 & 15.10 & 8.88 & 14.33 \\
    \hspace{1em} 2 & 3497  & 500  & 65.5 & 10.97$_{\textcolor{teal}{-9.6\%}}$ & 13.15$_{\textcolor{teal}{-12.9\%}}$ & 6.99$_{\textcolor{teal}{-21.3\%}}$ & 11.93$_{\textcolor{teal}{-16.7\%}}$ \\
    \hspace{1em} 3 & 7219  & 2000 & 68.6 & 10.50$_{\textcolor{teal}{-4.3\%}}$ & 12.46$_{\textcolor{teal}{-5.2\%}}$ & 4.61$_{\textcolor{teal}{-34.0\%}}$ & 10.94$_{\textcolor{teal}{-8.3\%}}$ \\
    \hspace{1em} 4 & 10262 & 2000 & 151.9 & 10.45$_{\textcolor{teal}{-0.5\%}}$ & 12.46$_{\textcolor{teal}{-0.0\%}}$ & 4.15$_{\textcolor{teal}{-10.0\%}}$ & 10.54$_{\textcolor{teal}{-3.7\%}}$ \\ \hline
    \textbf{Indonesian} \\
    \hspace{1em} 1 & 5765 & 2000 & 68.6 & 16.68 & 15.99 & 19.82 & 16.29 \\
    \hspace{1em} 2 & 4534 & 2000 & 68.6 & 15.60$_{\textcolor{teal}{-6.5\%}}$ & 15.23$_{\textcolor{teal}{-4.8\%}}$ & 15.83$_{\textcolor{teal}{-20.1\%}}$ & 14.30$_{\textcolor{teal}{-12.2\%}}$ \\
    \hspace{1em} 3 & 5714 & 2000 & 151.9  & 14.58$_{\textcolor{teal}{-6.5\%}}$ & 14.92$_{\textcolor{teal}{-2.0\%}}$ & 13.83$_{\textcolor{teal}{-12.6\%}}$ & 13.77$_{\textcolor{teal}{-3.7\%}}$ \\ \hline
    \textbf{Vietnamese} \\
    \hspace{1em} 1 & 2351 & 2000 & 68.6 & 16.08 & 16.95 & 24.63 &  17.86 \\
    \hspace{1em} 2 & 1764 & 2000 & 68.6 & 15.08$_{\textcolor{teal}{-6.2\%}}$ & 14.72$_{\textcolor{teal}{-13.2\%}}$ & 18.81$_{\textcolor{teal}{-23.6\%}}$ & 13.50$_{\textcolor{teal}{-24.4\%}}$ \\
    \hspace{1em} 3 & 6039 & 2000 & 151.9 & 14.09$_{\textcolor{teal}{-6.6\%}}$ & 12.83$_{\textcolor{teal}{-12.8\%}}$ & 14.43$_{\textcolor{teal}{-23.3\%}}$ & 11.59$_{\textcolor{teal}{-14.1\%}}$ \\
\bottomrule[1pt]
\end{tabular}}
\end{table*}
Table~\ref{tab:performance_nst} presents the ASR results across different NST iterations on three evaluation sets, including the development and test sets from GigaSpeech 2 and the Common Voice 17.0 and FLEURS test set.
Each iteration involves distinct modifications aimed at refining high-quality transcriptions.
A subset of automatic transcriptions generated by Whisper large-v3 is used to train the initial teacher model (Iter.~1). The teacher model then filters the training utterances by applying a CER/WER threshold, using the original labels as references and the new labels generated by the teacher as the hypothesis. The student model is trained on this filtered set with noise injected (Iter.~2). The student model is then used as the teacher to generate new labels on a larger subset of raw automatic transcriptions, applying the same filter to refine the training data. This refined data is used to train the student model with noise injected (Iter.~3). The process repeats in subsequent iterations, and the model size is scaled up to a larger version in the final iteration (Iter.~3 of Indonesian \& Vietnamese, Iter.~4 of Thai).

According to the results shown in Table~\ref{tab:performance_nst}, several noTable~trends can be observed:

1) Across all three languages, iteratively scaling the training data size, adding noise, and filtering labels lead to consistent improvements in the WER performance on the evaluation sets until the final iteration. This indicates that the iterative approach of refining and scaling the training data is effective in enhancing the accuracy of the raw transcriptions.

2) The system trained on Thai consistently achieves the absolute lowest error rates consistently across iterations from 1 to 4, indicating the effectiveness of the NST approach for this particular language. The best NST model outperforms the standard transcription model data by WER reductions of 1.69\%, 2.64\%, 4.73\%, and 3.79\% absolute (13.92\%, 17.48\%, 53.27\%, and 26.45\% relative) respectively (Iter.~4 \textit{vs.} 1).

Additional ablation studies on our modified NST in Appendix~\ref{sec:ablation_nst} Table~\ref{tab:ablation_nst} demonstrate the effectiveness of relabeling and discuss the detriment of enlarging noise when scaling the training data.

\begin{table*}[t]
  \small
  \centering
  \caption{Comparison of ASR results for models trained on GigaSpeech 2 with open-source multilingual ASR models and commercial ASR services, evaluated on three test sets from GigaSpeech 2, Common Voice 17.0, and FLEURS. The evaluation metrics include CER for Thai and WER for both Indonesian and Vietnamese. ``$\dagger$" denotes commercial services.}
  \label{tab:benchmarks}
  \renewcommand{\arraystretch}{1.15}
  \renewcommand\tabcolsep{12pt}
  \resizebox{0.95\linewidth}{!}{
    \begin{tabular}{lccccc}
    \toprule[1pt]
    \multirow{2}{*}{\textbf{Model}} & \multirow{2}{*}{\makecell[c]{\textbf{\#Params} \\ \textbf{(M)}}} & \multicolumn{3}{c}{\textbf{CER / WER}} \\
    & & GigaSpeech 2 & Common Voice & FLEURS \\ \hline
    \textbf{Thai} \\
    \hspace{1em} Whisper large-v3                  & 1542  & 20.44 & 6.02  & 11.55 \\
    \hspace{1em} Whisper large-v2                  & 1541  & 22.47 & 8.79  & 15.50 \\
    \hspace{1em} Whisper base                      & 72    & 46.47 & 32.59 & 42.28 \\ 
    \hspace{1em} MMS L1107                         & 964   & 31.75 & 14.49 & 23.07 \\
    \hspace{1em} Azure Speech CLI $1.37.0^\dagger$ & -     & 17.25 & 10.20 & 13.35 \\ 
    \hspace{1em} Google USM Chirp v$2^\dagger$     & -     & 49.70 & 14.75 & 63.35 \\ 
    \hspace{1em} GigaSpeech 2 (proposed)           & 151.9 & \textbf{12.46} & \textbf{4.15} & \textbf{10.54} \\ \hline
    \textbf{Indonesian} \\
    \hspace{1em} Whisper large-v3                  & 1542  & 20.03 & 7.43  & 7.85 \\
    \hspace{1em} Whisper large-v2                  & 1541  & 21.44 & 8.93  & 8.95 \\
    \hspace{1em} Whisper base                      & 72    & 39.37 & 34.70 & 33.76 \\ 
    \hspace{1em} MMS L1107                         & 964   & 35.27 & 20.72 & 24.49 \\
    \hspace{1em} Azure Speech CLI $1.37.0^\dagger$ & -     & 18.07 & 10.33 & 11.18 \\ 
    \hspace{1em} Google USM Chirp v$2^\dagger$     & -     & 19.63 & 9.70  & \textbf{7.23} \\ 
    \hspace{1em} GigaSpeech 2 (proposed)           & 151.9 & \textbf{14.92} & 13.83 & 13.77 \\ 
    \hspace{2em} + Common Voice + FLEURS           & 151.9 & 14.95 & \textbf{7.33} & 12.74 \\ \hline
    \textbf{Vietnamese} \\
    \hspace{1em} Whisper large-v3                  & 1542  & 17.94 & 13.74 & \textbf{8.59} \\
    \hspace{1em} Whisper large-v2                  & 1541  & 18.74 & 18.00 & 10.26 \\
    \hspace{1em} Whisper base                      & 72    & 39.88 & 44.07 & 40.41 \\ 
    \hspace{1em} MMS L1107                         & 964   & 46.62 & 43.88 & 55.35 \\
    \hspace{1em} Azure Speech CLI $1.37.0^\dagger$ & -     & \textbf{11.86} & \textbf{10.21} & 11.88 \\ 
    \hspace{1em} Google USM Chirp v$2^\dagger$     & -     & 13.28 & 12.46 & 11.75 \\ 
    \hspace{1em} GigaSpeech 2 (proposed)           & 151.9 & 12.83 & 14.43 & 11.59 \\
    \hspace{2em} + Common Voice + FLEURS           & 151.9 & 12.39 & 11.47 & 9.94 \\
\bottomrule[1pt]
\end{tabular}}
\end{table*}
\begin{table*}[t]
  \centering
  \small
  \caption{Comparison of ASR results for models trained on YODAS and GigaSpeech 2, evaluated on test sets from GigaSpeech 2, Common Voice 17.0, and FLEURS. The evaluation metrics include CER for Thai and WER for both Indonesian and Vietnamese.}
  \label{tab:yodas}
  \renewcommand{\arraystretch}{1.15}
  \renewcommand\tabcolsep{15pt}
  \resizebox{0.95\linewidth}{!}{
    \begin{tabular}{lccccc}
    \toprule[1pt]
    \multirow{2}{*}{\textbf{Training Set}} & \multirow{2}{*}{\makecell[c]{\textbf{\#Params} \\ \textbf{(M)}}} & \multicolumn{3}{c}{\textbf{CER / WER}} \\
    & & GigaSpeech 2 & Common Voice & FLEURS \\ \hline
    \textbf{Thai} \\
    \hspace{1em} YODAS manual                  & 68.6  & 27.34 & 10.71 & 14.19 \\
    \hspace{1em} YODAS manual                  & 151.9 & 28.76 & 10.96 & 16.11 \\
    \hspace{1em} \textit{GigaSpeech 2 refined} & 151.9 & \textbf{12.46} & \textbf{4.15} & \textbf{10.54} \\ \hline
    \textbf{Indonesian} \\
    \hspace{1em} YODAS manual                  & 68.6  & 25.77 & \textbf{10.82} & 14.63 \\
    \hspace{1em} YODAS manual + automatic      & 68.8  & 41.11 & 15.41 & 47.26 \\
    \hspace{1em} YODAS manual                  & 151.9 & 25.11 & 11.05 & \textbf{12.67} \\
    \hspace{1em} \textit{GigaSpeech 2 refined} & 151.9 & \textbf{14.92} & 13.83 & 13.77 \\ \hline
    \textbf{Vietnamese} \\
    \hspace{1em} YODAS manual                  & 68.6  & 40.35 & 31.07 & 25.68 \\
    \hspace{1em} YODAS manual + automatic      & 68.6  & 71.91 & 25.73 & 61.38 \\
    \hspace{1em} YODAS manual                  & 151.9 & 40.71 & 32.58 & 29.32 \\
    \hspace{1em} \textit{GigaSpeech 2 refined} & 151.9 & \textbf{12.83} & \textbf{14.43} & \textbf{11.59} \\
\bottomrule[1pt]
\end{tabular}}
\end{table*}
\begin{table*}[t]
  \small
  \centering
  \caption{Comparison of ASR models trained on GigaSpeech 2 with Icefall and ESPnet toolkits, evaluated on GigaSpeech 2 TEST set. The evaluation metrics include CER for Thai (th) and WER for both Indonesian (id) and Vietnamese (vi).}
  \label{tab:baseline}
  \renewcommand{\arraystretch}{1.15}
  \resizebox{0.95\linewidth}{!}{
  \renewcommand\tabcolsep{15.0pt}
    \begin{tabular}{cccccc}
    \toprule[1pt]
    \multirow{2}{*}{\textbf{Toolkit}} & \multirow{2}{*}{\textbf{Model}} & \multirow{2}{*}{\makecell[c]{\textbf{\#Params} \\ \textbf{(M)}}} & \multicolumn{3}{c}{\textbf{CER / WER}} \\
    & & & th & id & vi \\ \hline
    Icefall & Zipformer/Stateless Pruned RNN-T & 151.9 & 12.46 & 14.92 & 12.83 \\
    ESPnet  & Conformer/Transformer CTC/AED    & 111.8 & 13.70 & 15.50 & 14.60 \\
\bottomrule[1pt]
\end{tabular}
}
\end{table*}

\subsection{Comparison to Existing ASR Systems}
To demonstrate the efficacy of our ASR models trained on GigaSpeech 2, several mainstream and competitive ASR systems, including Whisper~\cite{whisper} from OpenAI, MMS~\cite{mms} from Meta, and commercial services from Azure and Google, are used as benchmarks.

\noindent\textbf{Whisper:}
Our work builds upon Whisper~\cite{whisper}, a suite of large-scale, multitask, and multilingual speech models developed by OpenAI. It leverages the encoder-decoder Transformer architecture~\cite{transformer}, with model sizes ranging from 39 million parameters (tiny) to 1.55 billion parameters (large). Additionally, Whisper offers variants spanning from an English-only version to a multilingual model capable of handling 99 languages. To conduct a comprehensive evaluation, we test three variants: Whisper base, Whisper large-v2, and Whisper large-v3 models.

\noindent\textbf{MMS:}
The Massively Multilingual Speech (MMS)~\cite{mms} project leverages self-supervised learning (SSL) techniques and a novel dataset to expand the language coverage of speech technology significantly. The core components include pre-trained wav2vec 2.0 \cite{wav2vec2} models for 1,406 languages, a single multilingual ASR model supporting 1,107 languages, speech synthesis models for the same set of languages, and a language identification model capable of recognizing 4,017 languages. In this study, we employ the MMS L1107 configuration.

\noindent\textbf{Azure AI Speech:}
Azure Speech CLI offers a convenient way to leverage Microsoft's speech recognition capabilities directly from the command line. It not only supports a wide range of audio file formats but also possesses the ability to handle various streaming audio inputs. 
We utilize the Azure Speech CLI version 1.37 in this paper, which is the latest version available.

\noindent\textbf{Google USM:}
The Universal Speech Model (USM) \cite{usm} is introduced as a single, large-scale model that excels in ASR across over 100 languages. This achievement is made possible by pre-training the model's encoder on a vast, unlabeled multilingual dataset of 12 million hours, covering more than 300 languages, followed by fine-tuning on a smaller labeled dataset. To conduct a thorough comparison, we utilize their Chirp Speech-to-Text v2 model for performance evaluation.

We compare the performance of our proposed approach trained on GigaSpeech 2 against these above-mentioned ASR models, including Whisper (base, large-v2, and large-v3), MMS L1107, Azure Speech CLI 1.37.0 and Google USM Chirp v2\footnote{Abnormal high deletion rates with Google USM in Thai are observed in our repeated testing.}, across three languages: Thai, Indonesian, and Vietnamese. The ASR performance is evaluated regarding character error rate (CER) or word error rate (WER) on three distinct test sets from GigaSpeech 2, Common Voice 17.0, and FLEURS. According to the results shown in Table~\ref{tab:benchmarks}, there are several intriguing findings: 

\noindent 1) For the Thai language, our ASR model trained on GigaSpeech 2 (Table~\ref{tab:benchmarks}, Thai, Row 7) outperforms all competitors, including commercial services from Azure and Google, securing the top rank across all three test sets among the seven models.
It outperforms Whisper large-v3 by relative WER reductions of 39.04\%, 31.06\%, and 8.74\% (Table~\ref{tab:benchmarks}, Thai, Row 7 \textit{vs.} 1). Remarkably, our model achieves such impressive performance with nearly one-tenth of the parameters compared to Whisper large-v3 (151.9 M \textit{vs.} 1542 M). 

\noindent 2) For the Indonesian and Vietnamese languages, our system demonstrates competitive performance compared to existing baseline models. This highlights the efficacy of our pipeline in delivering high-quality results with a lightweight model.
Specifically, on the GigaSpeech 2 test set in the Indonesian language, our system (Table~\ref{tab:benchmarks}, Indonesian, Row 7) outperforms all baseline models, attaining the best performance.
Compared to Whisper large-v3, the model trained on Indonesian achieves an absolute WER reduction of 5.11\%, corresponding to a relative reduction of 25.51\% (Table~\ref{tab:benchmarks}, Indonesian, Row 7 \textit{vs.} 1). 
Similarly, the model trained on Vietnamese achieves an absolute WER reduction of 5.11\%, corresponding to a relative reduction of 28.48\% (Table~\ref{tab:benchmarks}, Vietnamese, Row 7 \textit{vs.} 1).

\noindent 3) Our model exhibits degraded performance compared to commercial ASR systems on the Common Voice and FLEURS test sets in Indonesian and Vietnamese, which can be attributed to the domain mismatch\footnote{Unlike GigaSpeech 2, which contains noisy, reverberant spontaneous speech, Common Voice and FLEURS comprise clean, read speech with text from written materials.}. Contrastively, we observe a performance leap after adding Common Voice and FLEURS training data into GigaSpeech 2 (Table~\ref{tab:benchmarks}, Indonesian \& Vietnamese, Row 7 \textit{vs.} 8).

Although our training data size is smaller than that of industrial-scale models, our method achieves the best performance for the Thai language domain and delivers comparable results to commercial models for Indonesian and Vietnamese. This remarkable accomplishment highlights the efficacy of our approach in leveraging limited, free, open-source, unlabeled data to train highly competitive speech recognition models. It showcases a promising path towards developing high-quality speech recognition systems without the need for extensive, proprietary datasets, thereby reducing the barrier to entry and enabling wider accessibility.

\subsection{Comparison to the YODAS Corpus}

Table~\ref{tab:yodas} compares ASR performance across different models trained on YODAS~\cite{yodas} and GigaSpeech 2 datasets evaluated on multiple test sets. Note that YODAS Thai automatic is not included due to insufficient data (only 1 hour). Despite variations in overall data volume, several general conclusions can be drawn from trend analysis:

\noindent 1) The models trained on \textit{GigaSpeech 2 refined} yield generally superior results compared to those trained on YODAS datasets for all three languages.

\noindent 2) The YODAS manual may suffer from overfitting or noisy data issues due to simplistic filtering rules, leading to inconsistent performance in Indonesian (Table~\ref{tab:yodas}, Indonesian, Row 1 \& 3).

\noindent 3) Purely automatic generation of YODAS tends to degrade performance, as observed for Vietnamese (Table~\ref{tab:yodas}, Vietnamese, Row 1 \textit{vs.} 2) and Indonesian (Table~\ref{tab:yodas}, Indonesian, Row 1 \textit{vs.} 2), likely due to the inherent noise and errors in the automatically generated subtitles.

\subsection{Training ASR Models within ESPnet and icefall on GigaSpeech 2}
\noindent\textbf{Icefall:}
We adopt the neural Transducer~\cite{rnnt} architecture, using Zipformer-L as the encoder, the pruned RNN-T loss~\cite{pruned-rnnt} as the object function, and 2000-class Byte Pair Encoding (BPE)~\cite{bpe} word pieces. More details are provided in Appendix~\ref{sec:config_zipformer}.

\noindent\textbf{ESPnet:}
We employ Conformer~\cite{conformer} CTC/AED~\cite{ctc/aed} system from ESPnet \cite{ESPnet}, using Conformer-L as the encoder and 2000-class BPE word pieces. This model combines the localized sensitivity of convolutional neural networks with the long-range modeling capabilities of Transformers \cite{transformer}. Details are available in Appendix~\ref{sec:config_conformer}.

Table~\ref{tab:baseline} shows the results of ASR models trained with icefall and ESPnet. The models trained with ESPnet are slightly worse than icefall in all three languages, which is as expected and can be explained by the discrepancy in the number of model parameters (112M \textit{vs.} 152M). It is worth noting that the results in Table~\ref{tab:baseline} are intended to provide baseline systems for these two popular toolkits to demonstrate the universality of GigaSpeech 2 instead of pursuing state-of-the-art performance.
\section{Conclusion}
This paper introduces a new multilingual speech dataset, GigaSpeech 2, and a novel automated pipeline to boost speech recognition performance using in-the-wild audio-only data. GigaSpeech 2 aims to address the scarcity of labeled training data on low-resource languages by developing this large-scale, multi-domain, and multilingual corpus. Extensive experiments are conducted to validate the efficacy of our newly introduced corpus. The ASR models trained in three languages, which are Thai, Indonesian, and Vietnamese within GigaSpeech 2, demonstrate superior and impressive performance compared to various powerful ASR models, including Whisper large v2/v3 from OpenAI, MMS from Meta, and even commercial services from Google and Azure. The related resources, including the corpus with curated test sets\footnote{\url{https://huggingface.co/datasets/speechcolab/gigaspeech2}}, automated pipeline\footnote{\url{https://github.com/SpeechColab/GigaSpeech2}}, and recipes\footnote{\url{https://github.com/k2-fsa/icefall}}, are released to facilitate research in this direction. In the future, we are eager to extend our paradigm to more low-resource languages and are devoted to breaking down the language barrier.

\section*{Limitations}
\label{sec:limitation}
In this paper, we propose GigaSpeech 2, a large-scale, multi-domain, multilingual speech recognition corpus, and a novel automated pipeline to boost speech recognition performance using in-the-wild audio-only data.
We only conducted 3-4 iterations of the proposed NST method in our experiments, and we are optimistic that more iterations on large data will yield even better results. 
Moreover, we are actively extending our language coverage by incorporating additional languages, including Malay, Korean, Minnan, and Arabic. We will also expand our low-resource language family in our future investigation.
\section*{Ethics Statement}
All collected audio is sourced from materials released under a Creative Commons license.
Personally identifiable information has been anonymized using rule-based scripts to remove identifiable content from the data.
All annotators are compensated fairly by a professional data annotation company.
Our dataset adopts the same terms as GigaSpeech~\cite{gigaspeech} to resolve potential legal risks, restricting use to non-commercial research and educational purposes only.
We are committed to ongoing maintenance of the dataset to address any potential risks in the future.
\section*{Acknowledgement}
This work was supported by the National Natural Science Foundation of China  (No. U23B2018 and No. 62206171), Shanghai Municipal Science and Technology Major Project under Grant 2021SHZDZX0102 and Yangtze River Delta Science and Technology Innovation Community Joint Research Project (2024CSJGG01100).
We gratefully acknowledge the support of DataOcean AI for manually annotating the evaluation sets.

\clearpage

\bibliography{custom}

\appendix
\section{Detailed Analysis of GigaSpeech 2}
\label{sec:detailed_analysis_gigaspeech2}

\subsection{Manual Transcription Quality Assurance}
The manual transcription process, carried out by a professional data annotation company, includes rigorous manual quality checks and secondary inspections to ensure that timestamp accuracy and transcription correctness exceed 97\%.
All manually transcribed results undergo a 100\% manual quality inspection, where both timestamps and transcription accuracy are thoroughly checked. Any data that fails to meet the required standards is sent back for correction. Subsequently, 30\% of each inspector’s reviewed data is re-evaluated. If this recheck confirms over 97\% accuracy, the data passes; otherwise, the entire dataset inspected by that quality inspector is returned for full correction.
For timestamp accuracy, an audio snippet tool is used to ensure that timestamps do not overlap with the waveform. If any timestamp does fall on the waveform, a manual inspection is conducted to confirm whether it corresponds to speech.

\subsection{Domain Distribution of Manual Evaluation Sets}
\begin{figure}[ht]
    \centering
    \begin{minipage}{0.32\linewidth}
        \subfigure[th] {
        \label{fig:th_domain}
        \includegraphics[width=\linewidth]{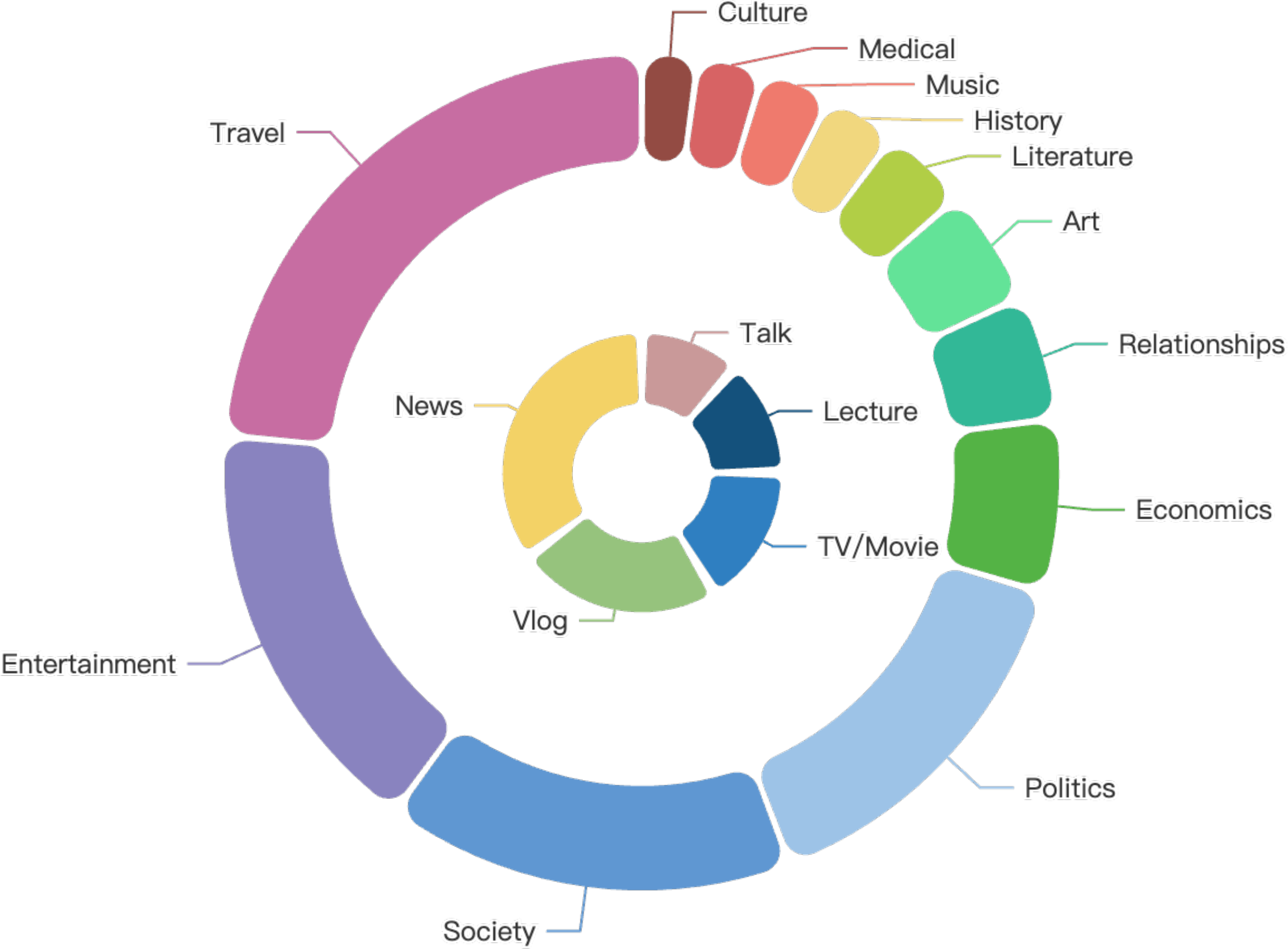}
        }
    \end{minipage}
    \begin{minipage}{0.32\linewidth}
        \subfigure[id] {
        \label{fig:id_domain}
        \includegraphics[width=\linewidth]{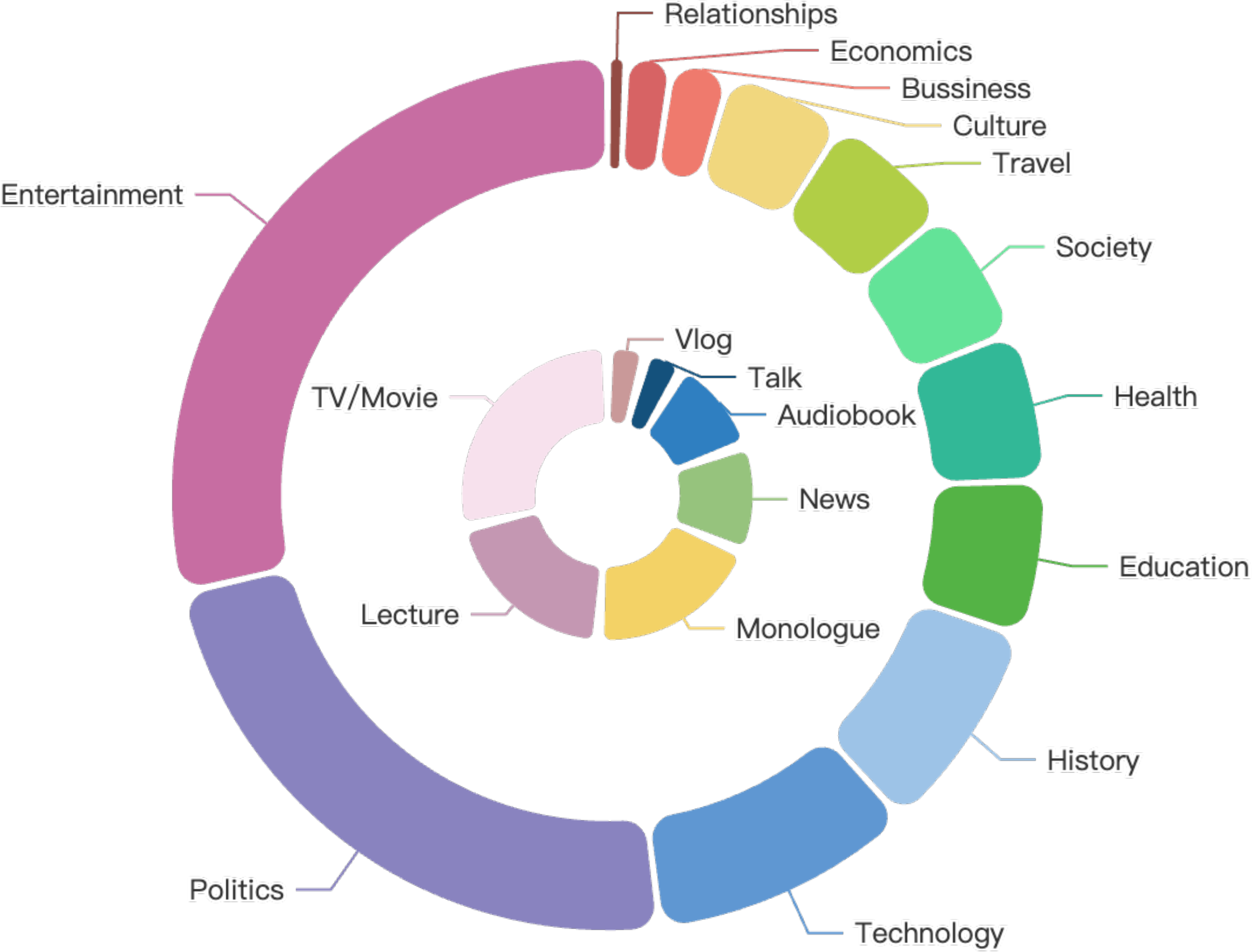}
        }
    \end{minipage}
    \begin{minipage}{0.32\linewidth}
        \subfigure[vi] {
        \label{fig:vi_domain}
        \includegraphics[width=\linewidth]{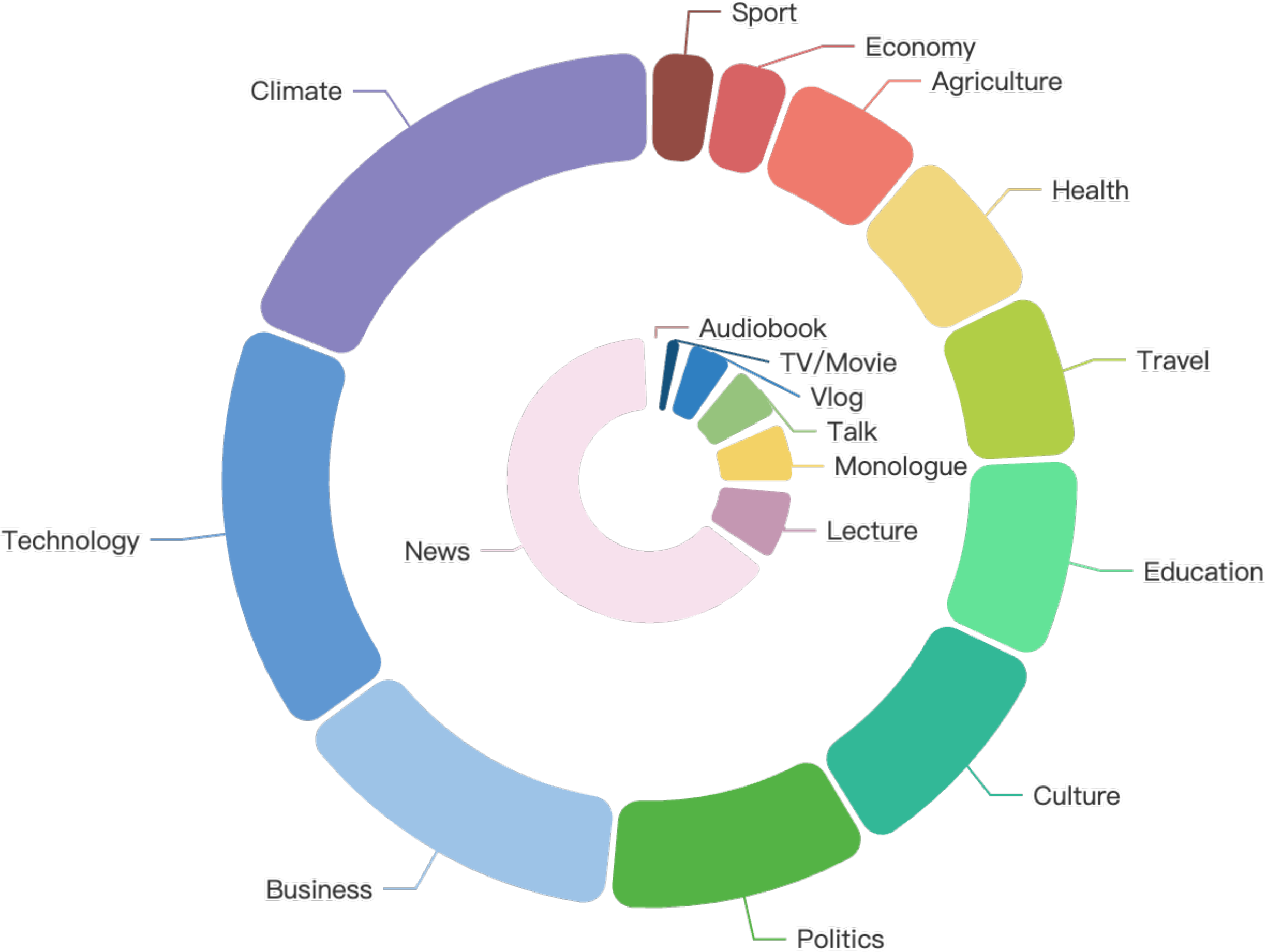}
        }
    \end{minipage}
    \caption{Hours distribution of manual evaluation sets for Thai (th), Indonesian (id), and Vietnamese (vi). The inner circle represents the format, and the outer circle represents the topic.}
    \label{fig:domain_distribution}
\end{figure}

The domain distribution of the manual evaluation sets is shown in Fig.~\ref{fig:domain_distribution}. The domains are identified based on a predefined set of categories. Each sample is manually annotated at the individual video level, considering both the topic type and content format.

\subsection{Duration Distribution of Training Sets}
\begin{figure}[ht]
    \centering
    \begin{minipage}{0.3\linewidth}
        \subfigure[th raw] {
        \label{fig:th_raw_duration}
        \includegraphics[width=\linewidth]{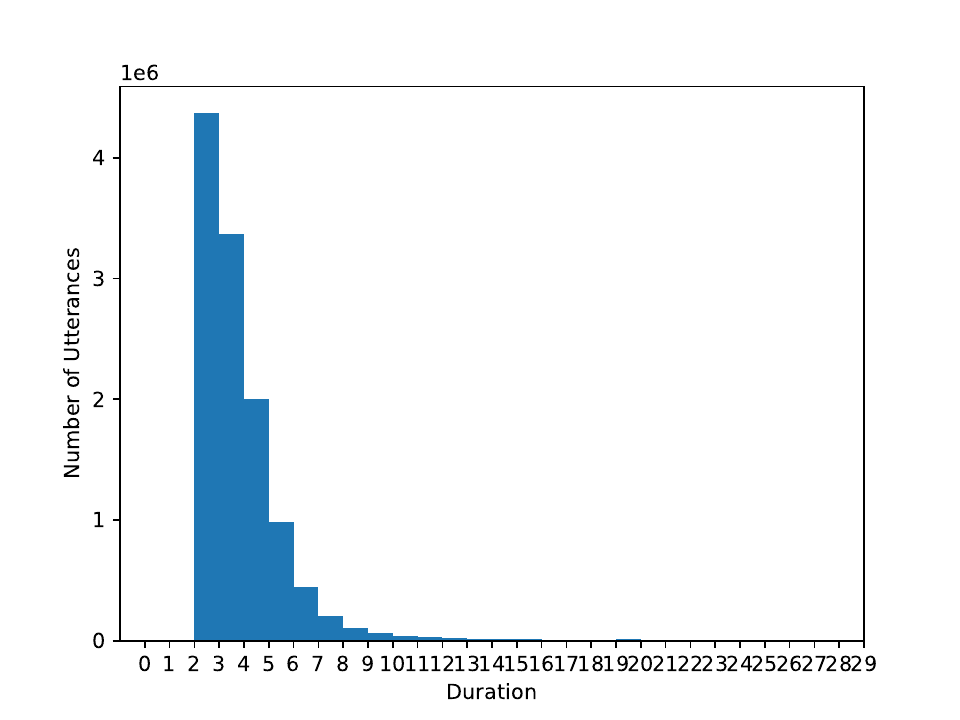}
        }
        \subfigure[th refined] {
        \label{fig:th_refined_duration}
        \includegraphics[width=\linewidth]{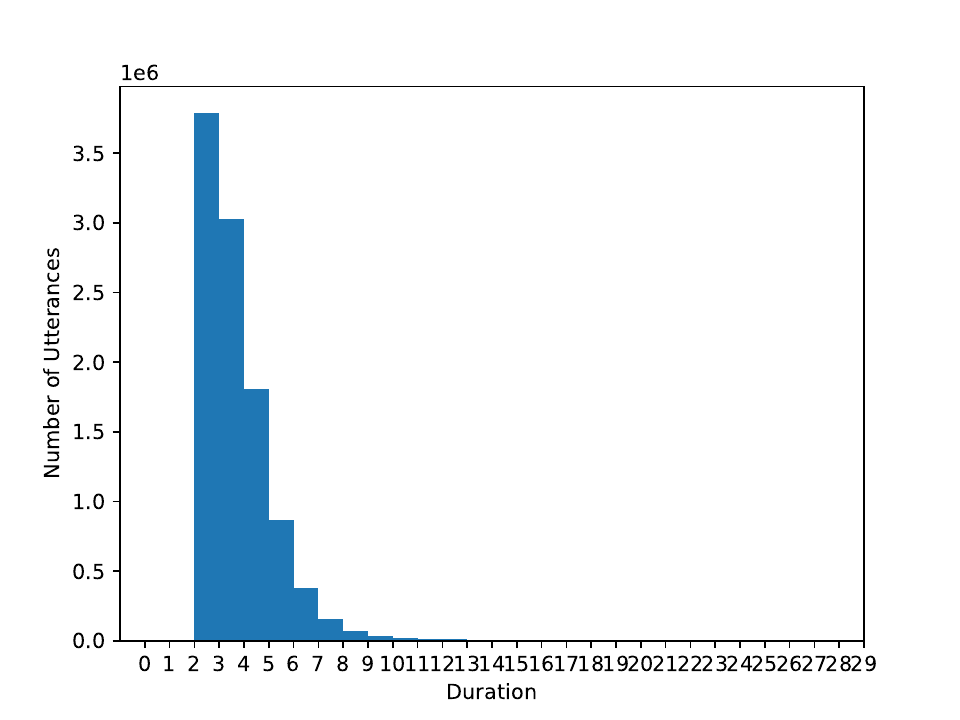}
        }
    \end{minipage}
    \begin{minipage}{0.3\linewidth}
        \subfigure[id raw] {
        \label{fig:id_raw_duration}
        \includegraphics[width=\linewidth]{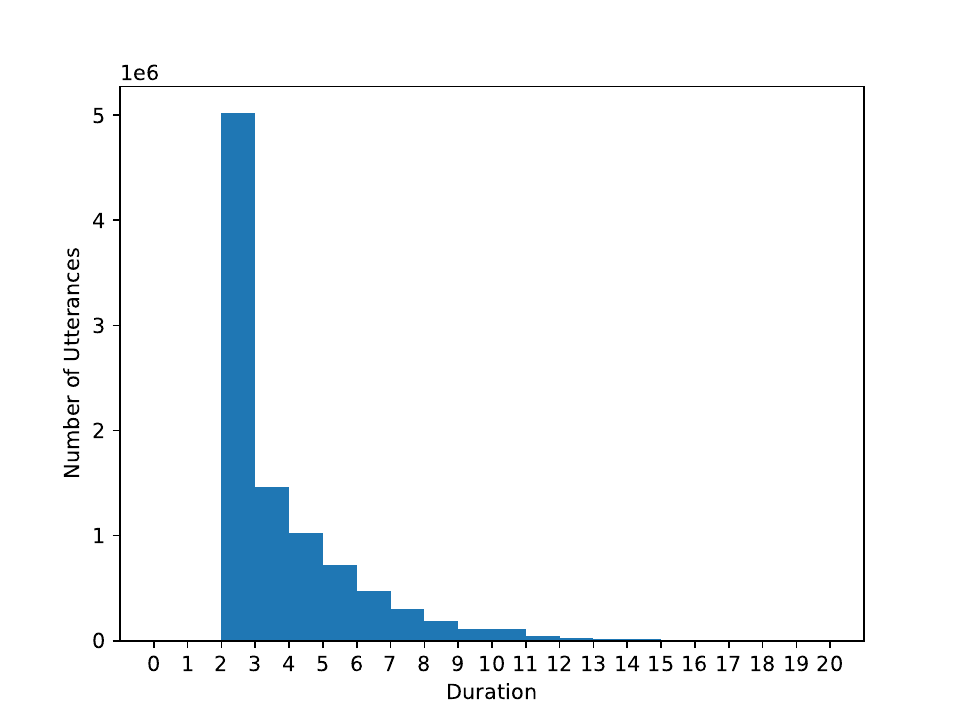}
        }
        \subfigure[id refined] {
        \label{fig:id_refined_duration}
        \includegraphics[width=\linewidth]{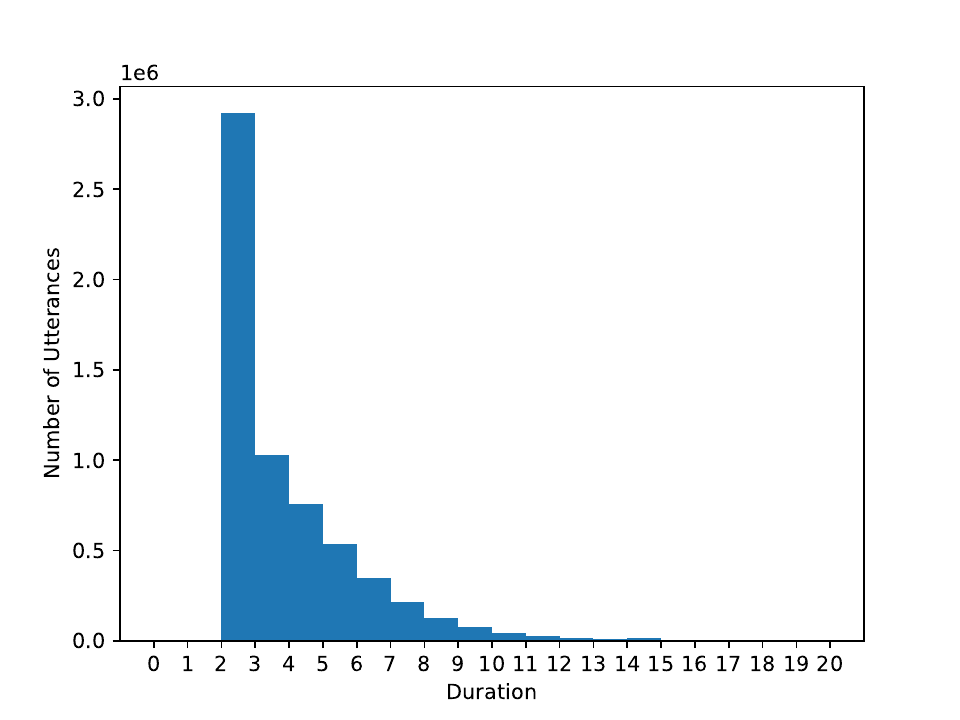}
        }
    \end{minipage}
    \begin{minipage}{0.3\linewidth}
        \subfigure[vi raw] {
        \label{fig:vi_raw_duration}
        \includegraphics[width=\linewidth]{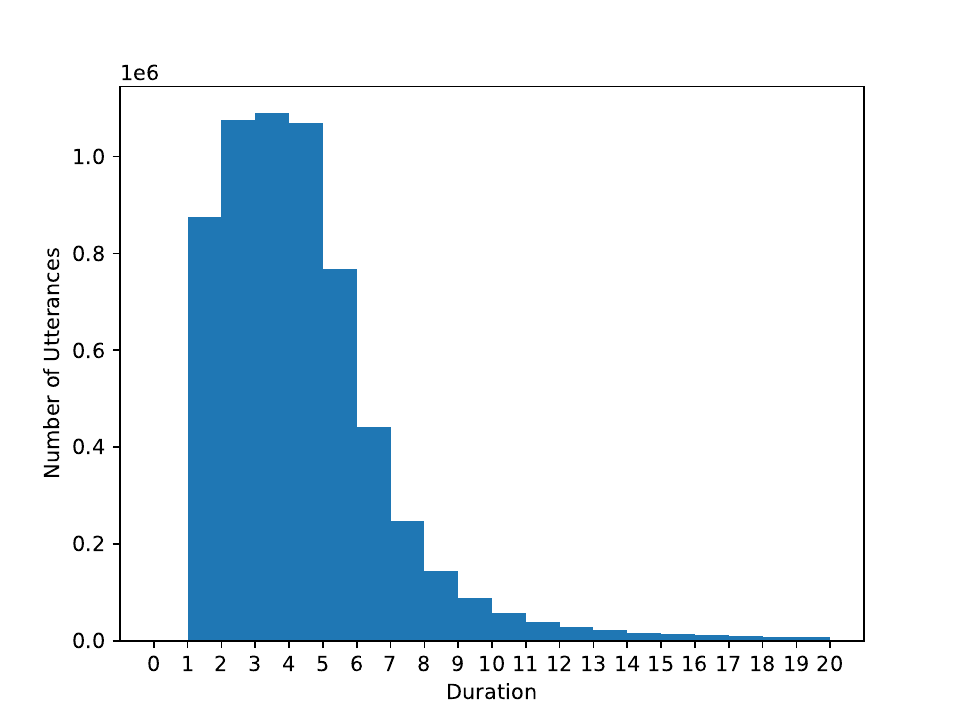}
        }
        \subfigure[vi refined] {
        \label{fig:vi_refined_duration}
        \includegraphics[width=\linewidth]{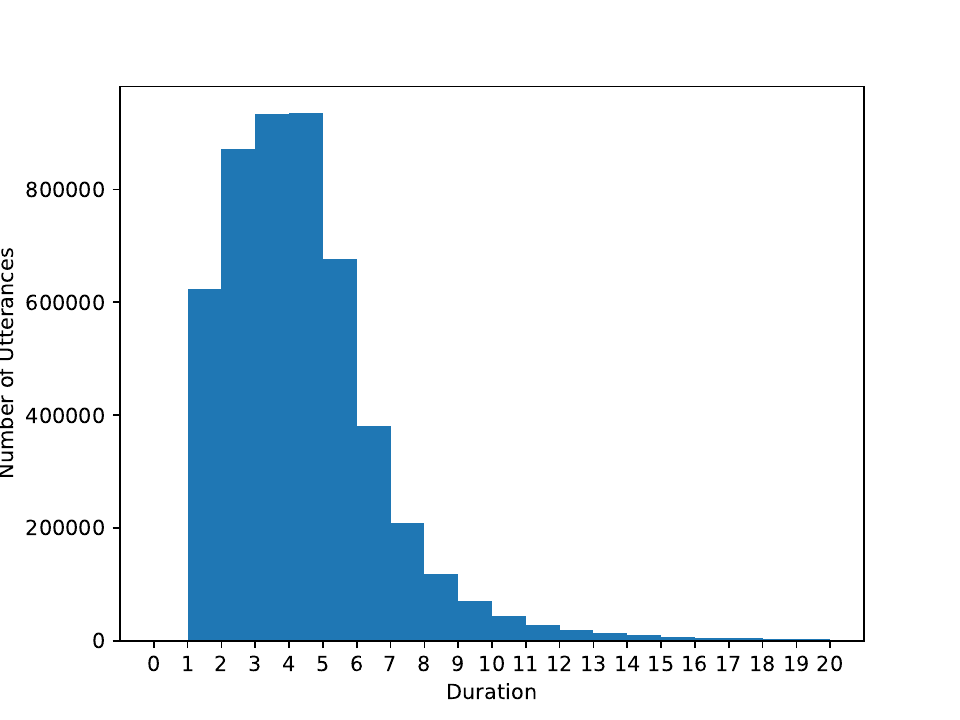}
        }
    \end{minipage}
    \caption{Utterance-level duration (second) distribution of training sets for Thai (th), Indonesian (id), and Vietnamese (vi).}
    \label{fig:duration_distribution}
\end{figure}

The utterance-level duration distribution of the training sets is illustrated in Fig.~\ref{fig:duration_distribution}.

\subsection{Evaluation of Processing Time}
The processing times for transcription, forced alignment, filtering, segmentation, and relabeling are measured on an idle single V100 32G GPU machine using a 100-hour subset of Thai audio. The processing time and the real-time factor (RTF) are detailed in Table~\ref{tab:evaluate_overall_time}.

\begin{table}[ht]
  \small
  \centering
  \caption{Evaluation of overall processing time and real-time factor (RTF) for each process in the construction of GigaSpeech 2. The processing times for transcription, forced alignment, filtering, segmentation, and relabeling are measured on an idle single V100 32G GPU machine using a 100-hour subset of Thai audio.}
  \label{tab:evaluate_overall_time}
  \renewcommand{\arraystretch}{1.1}
  \renewcommand\tabcolsep{10.0pt}
  \resizebox{\linewidth}{!}{
  \begin{tabular}{lll}
  \toprule[1pt]
  \textbf{Process} & \textbf{Time Consumption}  & \textbf{RTF} \\\hline
  Transcription    & 19h 42min 13s & $1.97 \times 10^{-1}$ \\
  Forced Alignment & 3h 27min 29s  & $3.46 \times 10^{-2}$ \\
  Filter           & 3s            & $8.00 \times 10^{-6}$ \\
  Segmentation     & 6min 58s      & $1.16 \times 10^{-3}$ \\
  Relabel          & 40min 48s     & $6.80 \times 10^{-3}$ \\
  \bottomrule[1pt]
  \end{tabular}}
\end{table}

\section{Model Configurations}
\begin{table}[ht]
  \small
  \centering
  \caption{Configuration of Zipformer at two different scales}
  \label{tab:config_zipformer}
  \renewcommand{\arraystretch}{1.1}
  \renewcommand\tabcolsep{0pt}
  \resizebox{\linewidth}{!}{
    \begin{tabular}{lcc}
    \toprule[1pt]
    & \textbf{Zipformer-M} & \textbf{Zipformer-L} \\\hline
    \textbf{Encoder} \\
    \hspace{1em} number of stacks                                 & \multicolumn{2}{c}{6} \\
    \hspace{1em} numbers of layers                                & {2,2,3,4,3,2}                & {2,2,4,5,4,2} \\
    \hspace{1em} downsampling factors                             & \multicolumn{2}{c}{1,2,4,8,4,2} \\
    \hspace{1em} output downsampling factor                       & \multicolumn{2}{c}{2} \\
    \hspace{1em} embedding dimensions                             & {192,256,384,512,384,256}    & {192,256,512,768,512,256} \\
    \hspace{1em} embedding unmasked dimensions                    & {192,192,256,256,256,192}    & {192,192,256,320,256,192} \\
    \hspace{1em} feedforward dimensions                           & {512,768,1024,1536,1024,768} & {512,768,1536,2048,1536,768}  \\
    \hspace{1em} convolution kernel sizes                         & \multicolumn{2}{c}{31,31,15,15,15,31} \\
    \hspace{1em} attention heads                                  & \multicolumn{2}{c}{4,4,4,8,4,4} \\
    \hspace{1em} attention query dimension                        & \multicolumn{2}{c}{32} \\
    \hspace{1em} attention value dimension                        & \multicolumn{2}{c}{12} \\
    \hspace{1em} positional encoding embedding dimension          & \multicolumn{2}{c}{48} \\
    \hspace{1em} projected positional encoding dimension per head & \multicolumn{2}{c}{4} \\
    \textbf{Decoder} \\
    \hspace{1em} embedding dimensions                             & \multicolumn{2}{c}{512} \\
    \hspace{1em} context size                                     & \multicolumn{2}{c}{2} \\
    \textbf{Joiner} \\
    \hspace{1em} embedding dimensions                             & \multicolumn{2}{c}{512} \\
    \textbf{Criterion} \\
    \hspace{1em} use ctc head                                     & \multicolumn{2}{c}{false} \\
    \hspace{1em} use transducer head                              & \multicolumn{2}{c}{true} \\
    \hspace{1em} pruned range                                     & \multicolumn{2}{c}{5} \\
    \hspace{1em} loss smoothing lm scale                          & \multicolumn{2}{c}{0.25} \\
    \hspace{1em} loss smoothing am scale                          & \multicolumn{2}{c}{0.0} \\
    \hspace{1em} simple loss scale                                & \multicolumn{2}{c}{0.5} \\
    \hspace{1em} simple loss scale warmup steps                   & \multicolumn{2}{c}{2000} \\
    \textbf{Frontend} \\
    \hspace{1em} n fft                                            & \multicolumn{2}{c}{512} \\
    \hspace{1em} hop length                                       & \multicolumn{2}{c}{256} \\
    \hspace{1em} feature dimension                                & \multicolumn{2}{c}{80} \\
    \textbf{Training} \\
    \hspace{1em} use amp                                          & \multicolumn{2}{c}{true} \\
    \hspace{1em} max epochs                                       & \multicolumn{2}{c}{30} \\
    \hspace{1em} max duration per batch                           & \multicolumn{2}{c}{1000} \\
    \hspace{1em} ref duration                                     & \multicolumn{2}{c}{600} \\
    \hspace{1em} seed                                             & \multicolumn{2}{c}{42} \\
    \textbf{Optimization} \\
    \hspace{1em} optimizer                                        & \multicolumn{2}{c}{scaledadam} \\
    \hspace{1em} base learning rate                               & \multicolumn{2}{c}{0.045} \\
    \hspace{1em} seed                                             & \multicolumn{2}{c}{42} \\
    \textbf{Scheduler} \\
    \hspace{1em} scheduler                                        & \multicolumn{2}{c}{eden} \\
    \hspace{1em} lr batches                                       & \multicolumn{2}{c}{7500} \\
    \hspace{1em} lr epochs                                        & \multicolumn{2}{c}{10000 / training set hours} \\
    \hspace{1em} warmup batches                                   & \multicolumn{2}{c}{500} \\
    \hspace{1em} warmup starting lr                               & \multicolumn{2}{c}{0.5} \\
    \textbf{SpecAugment} \\
    \hspace{1em} time warping factor                              & \multicolumn{2}{c}{80} \\
    \hspace{1em} number of time masks                             & \multicolumn{2}{c}{10} \\
    \hspace{1em} time mask maximum width                          & \multicolumn{2}{c}{100} \\
    \hspace{1em} number of frequency masks                        & \multicolumn{2}{c}{2} \\
    \hspace{1em} frequency mask width range                       & \multicolumn{2}{c}{0 - 27} \\
  \bottomrule[1pt]
  \end{tabular}}
\end{table}
\begin{table*}[t]
  \small
  \centering
  \caption{Ablation study of NST on GigaSpeech 2 Thai, evaluated across various evaluation sets, including GigaSpeech 2 DEV and TEST, Common Voice 17.0 TEST, and FLEURS TEST.}
  \label{tab:ablation_nst}
  \renewcommand{\arraystretch}{1.1}
  \renewcommand\tabcolsep{10pt}
  \resizebox{\linewidth}{!}{
    \begin{tabular}{lcccc}
    \toprule[1pt]
    \multirow{3}{*}{\makecell[c]{\textbf{NST} \\ \textbf{method}}} & \multicolumn{4}{c}{\textbf{CER}} \\
    & \multicolumn{2}{c}{\makecell[c]{GigaSpeech 2 \\ DEV \; TEST}} &  \makecell[c]{Common Voice \\ TEST} & \makecell[c]{FLEURS \\ TEST} \\ \hline
    Sys. 1 (Tab. \ref{tab:performance_nst}, iter 2 $\rightarrow$ iter 3)                      & 10.47 & 12.38 & 4.63 & 10.96 \\
    Sys. 2 (Tab. \ref{tab:performance_nst}, iter 2 $\rightarrow$ iter 3, without relabeling)  & 10.77$_{\textcolor{red}{+2.9\%}}$ & 12.90$_{\textcolor{red}{+4.2\%}}$ & 5.23$_{\textcolor{red}{+13.0\%}}$ & 10.72$_{\textcolor{teal}{-2.2\%}}$ \\
    Sys. 3 (Tab. \ref{tab:performance_nst}, iter 2 $\rightarrow$ iter 3, larger augmentation) & 10.65$_{\textcolor{red}{+1.7\%}}$ & 12.81$_{\textcolor{red}{+3.5\%}}$ & 5.36$_{\textcolor{red}{+15.8\%}}$ & 10.86$_{\textcolor{teal}{-0.9\%}}$ \\
\bottomrule[1pt]
\end{tabular}}
\end{table*}
\begin{table}[ht]
  \small
  \centering
  \caption{Configuration of Conformer at the large scale.}
  \label{tab:config_conformer}
  \renewcommand{\arraystretch}{1.1}
  \renewcommand\tabcolsep{0pt}
  \resizebox{\linewidth}{!}{
    \begin{tabular}{lclc}
    \toprule[1pt]
    \multicolumn{4}{c}{\textbf{Conformer-L}} \\\hline
    \textbf{Encoder} & & \textbf{Criterion} \\
    \hspace{2em} attention head                & 8             &\hspace{2em} ctc weight & 0.3  \\
    \hspace{2em} numbers of blocks             & 12            &\hspace{2em} label smoothing &0.1\\
    \hspace{2em} linear unit                   & 2048          &\hspace{2em} length normalized & false\\
    \hspace{2em} dropout rate                  & 0.1           &\textbf{Frontend } \\
    \hspace{2em} positional dropout rate       & 0.1           &\hspace{2em} n fft & 512\\
    \hspace{2em} attention dropout rate        & 0.1           &\hspace{2em} hop length & 256\\
    \hspace{2em} input layer                   & conv2d        &\textbf{Training }\\
    \hspace{2em} normalize before              & true          &\hspace{2em} use amp & true \\
    \hspace{2em} macaron style                 & true          &\hspace{2em} gradient accumulation & 4 \\
    \hspace{2em} relative position type        & latest        &\hspace{2em} max epochs & 20 \\
    \hspace{2em} position encoding layer       & rel\_pos      &\textbf{Optimization}\\
    \hspace{2em} self-attention layer          & rel\_selfattn &\hspace{2em} optimizer & adam\\
    \hspace{2em} activation type               & swish         &\hspace{2em} learning rate & 0.0025\\
    \hspace{2em} use cnn module                & true          &\hspace{2em} weight decay & 0.000001\\
    \hspace{2em} cnn module kernel             & 31            &\textbf{Scheduler}\\
    \textbf{Decoder}                           &               &\hspace{2em} scheduler & warmuplr \\
    \hspace{2em} attention heads               & 8             &\hspace{2em} warmup steps & 40000 \\
    \hspace{2em} linear units                  & 2048          &\textbf{SpecAugment}\\
    \hspace{2em} number of blocks              & 6             &\hspace{2em} time warp window & 5 \\
    \hspace{2em} dropout rate                  & 0.1           &\hspace{2em} frequency mask width range & 0 - 27\\
    \hspace{2em} positional dropout rate       & 0.1           &\hspace{2em} number of frequency masks & 2\\
    \hspace{2em} self-attention dropout rate   & 0.1           &\hspace{2em} time mask width ratio range & 0.0 - 0.05\\
    \hspace{2em} source attention dropout rate & 0.1           &\hspace{2em} number of time masks & 10 \\
  \bottomrule[1pt]
  \end{tabular}}
\end{table}
\label{sec:configuration}

\subsection{Configuration of Zipformer}
\label{sec:config_zipformer}
Two Zipformer-based models are used, following official configurations reported in icefall\footnote{\url{https://github.com/k2-fsa/icefall}}. In each Zipformer stack, the hidden dimensions of the first and last feedforward modules are 3/4 and 5/4 of the middle one, respectively. Ahead of the encoder, a convolution subsampling module with a stride of 2 reduces the frame rate to 50 Hz. The input consists of 80-channel FBank features extracted over windows of 25ms, strided by 10ms. The label decoder utilizes a stateless decoder~\cite{stateless}. 8 V100 32G GPUs are used for training.
Detailed configurations are provided in Table~\ref{tab:config_zipformer}.

\subsection{Configuration of Conformer}
\label{sec:config_conformer}
A Conformer-based model is developed adhering to the official configurations outlined in ESPnet\footnote{\url{https://github.com/ESPnet/ESPnet}}. The model comprises an encoder that employs the Conformer architecture and a decoder that leverages the Transformer architecture. Moreover, the parameters for both the encoder and decoder components, the optimization process, the scheduling mechanism, and SpecAugment settings are carefully designed to ensure a comprehensive and efficient model setup. 4 A100 80G GPUs are used for training.
The specifics of these configurations are detailed in Table~\ref{tab:config_conformer}.

\section{Ablation Study on Noisy Student Training}
\label{sec:ablation_nst}

Based on the ablation study of our proposed NST on the evaluation sets in Table~\ref{tab:ablation_nst}, we can analyze the effects of different iterations and their impact on performance:

\noindent 1) Relabeling the data during the transition from iteration 2 to 3 is crucial for improving performance (Sys.1 \textit{vs.} Sys.2).

\noindent 2) Larger augmentation applied in our NST process may hurt the performance (Sys.1 \textit{vs.} Sys.3).

These findings suggest that careful consideration of the relabeling and augmentation strategies is crucial for optimizing the performance of the NST model across different evaluation sets and domains.

\begin{table}[th]
\small
\centering
\caption{ASR performance of Whisper Medium with/without fine-tuning on GigaSpeech 2 Thai, tested on GigaSpeech 2 TEST and Common Voice 17.0 TEST}
\renewcommand{\arraystretch}{1.1}
\renewcommand\tabcolsep{5pt}
\resizebox{\linewidth}{!}{
\begin{tabular}{lcc}
\toprule
\multirow{2}{*}{\textbf{Model}} & \multicolumn{2}{c}{\textbf{CER}} \\ 
& GigaSpeech 2 & Common Voice \\
\midrule
Whisper medium & 37.55 & 16.41 \\
+ GigaSpeech 2 Thai fine-tuned & 14.15$_{\textcolor{teal}{-62.3\%}}$ & 6.92$_{\textcolor{teal}{-57.8\%}}$ \\
\bottomrule
\end{tabular}
}
\label{tab:whisper_thai_cer}
\end{table}
\section{Additional Results of Whisper Medium}
We evaluated Whisper medium\footnote{\url{https://huggingface.co/openai/whisper-medium}}  and its fine-tuned version\footnote{\url{https://huggingface.co/scb10x/monsoon-whisper-medium-gigaspeech2}} on GigaSpeech 2 Thai, using the test sets from GigaSpeech 2 and Common Voice 17.0. As shown in Table~\ref{tab:whisper_thai_cer}, fine-tuning resulted in an approximate 60\% relative CER reduction across two test sets, indicating the high quality of the GigaSpeech 2 Thai.

\end{document}